\documentclass[conference]{IEEEtran}
\IEEEoverridecommandlockouts
\usepackage{graphicx}
\usepackage{amsmath}
\usepackage{amsfonts}
\usepackage{mathrsfs}
\usepackage[OT1]{fontenc} 
\usepackage{enumerate}
\usepackage{amsthm}
\usepackage{tabularx}
\usepackage{arydshln}

\usepackage[colorlinks=true,linkcolor=blue,citecolor=blue]{hyperref}
\usepackage{mwe}
\usepackage{subfig}
\usepackage{float}
\usepackage{graphicx}
\usepackage{textcomp}
\usepackage{xcolor}
\usepackage{footnote}
\usepackage{makecell}
\usepackage{graphicx}
\usepackage{amsmath}
\usepackage{amsfonts}
\usepackage{mathrsfs}
\usepackage[OT1]{fontenc} 
\usepackage{enumerate}
\usepackage{amsthm}
\usepackage{tabularx}
\usepackage{arydshln}

\usepackage[draft]{todonotes}

\usepackage{algorithm}
\usepackage{algpseudocode}
\usepackage[inline]{enumitem}
\usepackage{mathtools}

\newcommand{\RNum}[1]{\uppercase\expandafter{\romannumeral #1\relax}}
\definecolor{bleudefrance}{rgb}{0.19, 0.55, 0.91}
\definecolor{chromeyellow}{rgb}{1.0, 0.65, 0.0}
\definecolor{asparagus}{rgb}{0.53, 0.66, 0.42}

\begin{document}
\title{Clustering Strategies of Cooperative Adaptive Cruise Control: Impacts on Human-driven Vehicles}

\author{\IEEEauthorblockN{Zijia Zhong \\ Mark Nejad \\ Earl E. Lee \RNum{2}}
\IEEEauthorblockA{Department of Civil and Environmental Engineering\\
University of Delaware\\
Newark, DE, USA\\
Email: \{zzhong, elee, nejad\}@udel.edu}
\and
\IEEEauthorblockN{Joyoung Lee}
\IEEEauthorblockA{John A. Reif, Jr. Department of Civil and \\Environmental Engineering\\
New Jersey Institute of Technology\\
Newark, NJ, USA\\
Email: jo.y.lee@njit.edu}
}

\maketitle
\begin{abstract}
As a promising application of connected and automated vehicles (CAVs), Cooperative Adaptive Cruise Control (CACC) is expected to be deployed on the public road in the near term. Thus far the majority of the CACC studies have been focusing on the overall network performance with limited insight on the potential impact of CAVs on human-driven vehicles (HVs).
This paper aims to quantify  the influence of CAVs on HVs by studying the high-resolution vehicle trajectory data that is obtained from microscopic simulation. Two clustering strategies for CACC are implemented: an ad hoc coordination one and a local coordination one. 
Results show that the local coordination outperforms the ad hoc coordination across all tested market penetration rates (MPRs) in terms of network throughput and productivity. The greatest performance difference between the two strategies is observed at 30\% and 40\% MPR for throughput and productivity, respectively. However, the distributions of the hard braking observations (as a potential safety impact) for HVs change significantly under local coordination strategy. Regardless of the clustering strategy, CAVs increase the average lane change frequency for HVs. 30\% MPR is the break-even point for local coordination, after which the average lane change frequency decreases from the peak 5.42 to 5.38. Such inverse relationship to MPR is not found in the ah hoc case and the average lane change frequency reaches the highest 5.48 at 40\% MPR.

\end{abstract}


\section{Introduction}
Cooperative adaptive cruise control (CACC) enables closely-coupled vehicular platoons by the extra layers of communication and automation. Being one of the most-studied application of CAVs,  CACC is expected to drastically increase mobility, decrease emission, while providing a safer and more convenient way for occupants. The CACC evaluation thus far has been focusing on the benefits that CAV could potentially bring to our transportation network.  The potential impact on non-equipped vehicles (i.e. Human-driven vehicles(HVs)) has been overlooked.

The motivation of the study is twofold. First, studying CACC under mixed traffic conditions in anticipation of its near-term deployment has gained an increasing amount of attention. It is the consensus that CACC can improve the performance of our transportation system. However, the possible change of the flow characteristic of HVs has often been overlooked. CACC may alter the behavior of the HVs especially when active clustering strategy is employed.  The expected impact on HVs from CACC could include:
\begin{enumerate*}[label=\roman*)]
\item the additional weaving during CACC clustering, 
\item the lane change by HVs induced by CACC clustering,
\item increase collision risk, and
\item lane blockage by CACC platoon.
\end{enumerate*}
Furthermore, the majority of CACC studies focused on the longitudinal movement and put less emphasis on the lateral movement of CAVs which is vital in cooperative driving. The lateral control when it comes to local coordination of forming CACC platoons has to be taken into account, as the development of CACC vehicle progresses. By implementing local coordination, the CACC platoon ratio and consequently the network performance can be increased.

In this paper, the comparison among ad hoc coordination and local coordination for CACC are made according to network-wide performance measures. The traffic flow characteristic of HVs  was investigated with the emphasis on the interaction between HVs and CAVs based on the high-resolution vehicle trajectory data extracted from microscopic simulation. The proposed methodology developed is also suitable for extracting the driving behavioral data for field deployment and modeling heterogeneous  traffic flow that is consisted of HVs and CAVs. Three scenarios are evaluated: 
\begin{enumerate*}[label=\roman*)]
\item base without CAVs,
\item CAVs with ad hoc coordination, and 
\item CAVs with local coordination algorithm (e.g., active platoon formation).
\end{enumerate*}

The organization of the remainder of the paper is as follows. Relevant research regarding research of CACC in mixed traffic will be reviewed in Section \ref{sect:Literature}, followed by the microscopic simulation framework in Section \ref{sect:framework}. The simulation results are presented and discussed in Section \ref{sect:result}. Lastly, findings and recommendations are discussed in Section \ref{sect:conclusion}.

\section{CACC Research in Mixed Traffic}
\label{sect:Literature}
\subsection{Impact of CACC}
CACC can positively increase the traffic performance with sufficient presence in the traffic flow. The reduced time headway and following distance have been recognized as the primary benefits of CACC.  Arnaout and Arnaout \cite{arnaout2014exploring} proposed the Flexible Agent-based Simulator of Traffic framework for evaluating CACC with ad hoc coordination. Moderate, saturated, and over-saturated demand scenarios were tested on a hypothetical four-lane high highway under various of CACC market penetration rates (MPRs).  The advantage of CACC showed when MPR was above 40\% and the network was able to serve 9,400 out of 10,000 vehicles per hour (vph).
Lee et al. \cite{lee2014mobility} evaluated the potential benefits for both mobility and safety under a wide range of traffic scenarios. It was found that the mobility benefits of CACC were shown at as less as 30\% MPR.

Songchitruksa et al. \cite{songchitruksa2016incorporating} evaluated the improvement of CACC on a 26-mile segment of the Dallas I-30 freeway. For simplification, zero demand from the on-ramps was assumed.  The highest throughput was observed as 4,400 vph with local coordination, where only rear-join to a platoon was allowed.
Van Arem et al. \cite{van2006impact} assessed the impact of CACC on freeway traffic flow on a 6-km, one-lane freeway with ramps distributed with 1.6-km interval. They found the capacity reached 4,250 vph per lane (vphpl) with full CACC penetration. With the same CACC model, Shladover et al. \cite{shladover2012impacts} studied the impact of CACC on freeway traffic flow on a one-lane freeway with no demand from ramps. The lane capacity was found to reach 3,600 vphpl at 90\% MPR of CACC. 

All of the above studies assessed the potential benefits brought by CACC on the overall network. However, none of them has investigated the impact that CACC could bring to HVs, especially under local coordination in which a free-agent CAV is actively seeking and subsequently performing lane change to join a platoon.

Nowakowski et al. \cite{nowakowski2010cooperative} studied the acceptance of the short following distance (ranging from 0.6 s to 1.1 s) enabled by CACC. As discovered, while all the drivers showed the willingness to accept the shorter following gaps, male participants were more likely to choose a shorter following distance.  The carry-over effect of the short headway in manual driving was exhibited even after the disengagement from the platoons in the KONVOI project \cite{casey1992changes}.
Gouy et al. \cite{GOUY2014264} investigated the behavioral adaptation effect that potentially caused by the short headway of CACC platoon using driving simulator. Participants were instructed to driver alongside with two CACC platoon configurations:
\begin{enumerate*}[label=\roman*)]
\item 10-truck platoon with 0.3 s intra-platoon headway, and 
\item 3-truck platoon with 1.4 s intra-platoon headway.
\end{enumerate*}
They found that smaller average time headway was observed when in the short headway scenario. In the first platoon scenarios, participants spent more time under a 1-second headway, which is deemed unsafe \cite{FAIRCLOUGH1997387}. 

Lin et al. \cite{LIN2009620} studied the time-gap of bus ACC system in three aspects:
\begin{enumerate*}[label=\roman*)]
\item preferred time-gaps for expressway driving,
\item time-gaps that maximized safety, and 
\item the influence of the secondary tasks to time-gaps. 
\end{enumerate*}
Calvert and Lint \cite{Calvert2017will} studied the negative effect of ACC on the system capacity using the propose ACC control. The simulation scenarios contained  various traffic flow conditions in terms of demand and flow composition (e.g., ACC vehicles, trucks, etc.). They concluded that the small negative effect on road capacity did exist and it was caused by the higher gap times of ACC.

\subsection{CACC Coordination}

There are three types of clustering strategies: ad hoc coordination, local coordination, and global coordination \cite{shladover2012impacts}. In this study, only the former two clustering strategies are evaluated due to the scope. Since global coordination requires advance planning for the travel demand at an origin-designation level. CAVs are coordinated to enter the highway in platoons. Global coordination is likely to be sub-optimal without a robust planning system to couple with the logistical challenge under dynamic traffic conditions.

\subsubsection{Ad hoc coordination}
Ad hoc coordination assumes random arrival of CAVs and no coordination among them. Therefore, the probability of driving behind another CAV is highly correlated to the market penetration rate (MPR). Ad hoc coordination has been observed in the majority of the research due to its simplicity in implementation and the lack of the conceptual framework for CAV platoon formation.  However, the ad hoc coordination is not likely to harness the full potential of CACC, as it does not fully utilize the short intra-platoon headway enabled by the CAV technology.  Exclusive lane, or other forms of managed lane, has been employed to aid the ad hoc clustering \cite{segata2012simulation}. In a sense, a CAV lane is a special form of coordination where the target lane is the CAV lane.

\subsubsection{Local coordination}
 
Local coordination facilities the platoon formation, where free-agent CAVs are actively seeking clustering opportunity in their surroundings.   The subject CAV, as well as the surrounding CAVs can be coordinated to change trajectory to facilitate clustering.  There are four basic types of lane change:  
\begin{enumerate*}[label = \roman*)]
\item free-agent-to-free-agent  lane change, 
\item free-agent-to-platoon lane change,
\item platoon-to-free-agent lane change, and 
\item platoon-to-platoon lane change \cite{wang2017developing}. 
\end{enumerate*}

Lee et al. \cite{lee2014mobility} developed a local coordination scheme which allows three ways to form a platoon:
\begin{enumerate*}[label = \roman*)]
\item front-join,
\item mid-join, and 
\item rear-join.
\end{enumerate*} 
The longitudinal control was a rule-based acceleration selection. Developed from Lee et al.'s coordination model, Zhong et al. \cite{Zhong2017a} implemented the MIXIC \cite{van2006impact} as car following model to study the CAV benefits for arterials. Lee et al.'s \cite{lee2014mobility} CACC algorithm was updated to using the EIDM for simulating CACC string behavior. The string formation and dispersion mechanism were enhanced, including preferential lane logic, platoon size restriction \cite{NAP25366}. Zhong \cite{zhong2018assessing} developed a CACC control model by combining E-IDM and the MOBIL model \cite{Kesting2007}. The MOBIL model is adopted as the mechanism to prevent lane changing of a free-agent CAV that may be potentially disruptive to the surrounding traffic. When a potential platooning opportunity is identified via V2V communication, the CACC system estimates the impacts on the immediate vehicles based on MOBIL should the lane change be initiated.  
The Lane-change Model with Relaxation and Synchronization (LMRS) model, proposed in \cite{Schakel2012}, gives a normalized strategic lane-change score by taking into consideration of route,  gain speed, and lane preference. For a higher desire score, the driver is willing to accept smaller headway and to decelerate more in LMRS. Calvert and Lint \cite{Calvert2017will} adopted the LMRS model in conjunction with the IDM+ \cite{Schakel2010effects} for evaluating the ACC. 

\section{Evaluation Framework}
\label{sect:framework}

\subsection{Simulation of Human Driving Behavior}
Human drivers can take into account more input stimulus (e.g., brake lights, next-nearest neighbors, etc.) with anticipation of the situation for the next few seconds \cite{treiber2013traffic}. All of these aspects can be formulated in terms of psycho-physiological models, such as Gipps' model \cite{Gipps1981A} and Weidemann model \cite{Wiedemann1974}. Weidemann model was re-calibrated in \cite{reiter1994empirical} using an instrumented vehicle to measure the thresholds among difference driving states. The Weidemann model is used by Vissim as the as the default car-following model. 

The Vissim car-following model also includes the tactical driving behavior, which carries certain planning in advance with a temporal horizon (multiple time steps) or spatial horizon that is beyond neighboring vehicles \cite{barcelo2010fundamentals}.
There are four different driving states in the Widemann model. 
\begin{enumerate*}[label=\roman*)]
\item free driving,
\item approaching,
\item following,and
\item braking.
\end{enumerate*} 
The acceleration is primarily determined by the current speed, speed difference, and gap to the preceding vehicle for each of the four driving state. The Wiedemann-99 model, suitable for freeway application, has ten calibratable parameters to represent a wide range of driver behavior. Therefore, the Wiedmann model has to be calibrated to specific traffic stream data \cite{higgs2011analysis, durrani2016calibrating},  as it was initially developed on limited available data. The objective of the calibration process is to minimize the difference between the measured driving behavior in the field and the driving behavior simulated.

There are two types of lane change in the Wiedemann model: necessary lane change and free lane change. The former focus on the hard constraint of the lane change, such as lane drop. The latter type is the focus of this study. Such lane change is performed when more space and higher speed is desired for a vehicle. As such, the safety distance plays an important role in the lane change behavior. A suitable gap is found based on
\begin{enumerate*}[label=\roman*)]
\item  the speed of the vehicle changing lane, and 
\item the approaching speed of the vehicle from behind on the lane \cite{ptv2018ptv}
\end{enumerate*} 

\subsection{Quantify Impact to Human Driver}
\begin{figure*} [h]
	\centering
	\includegraphics[width=\textwidth]{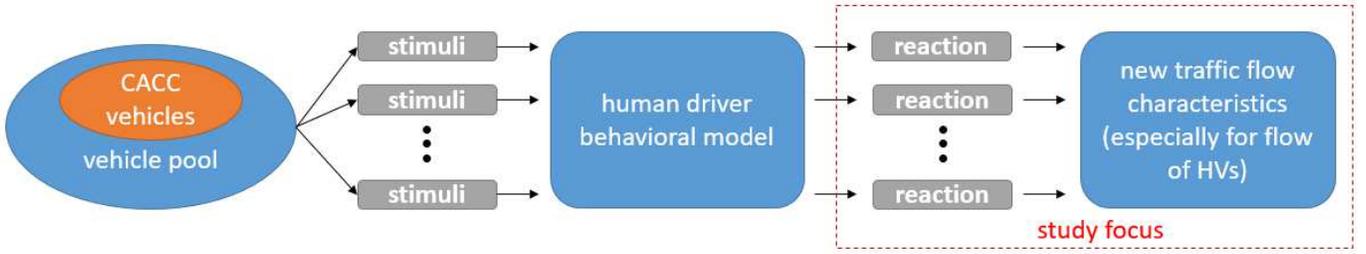}   
	\caption{Potential Impact to Non-equipped Vehicles} 
	\label{fig: impactEvaluationMethod}
\end{figure*}

The trajectory data for HVs are also collected. A before-and-after study is the most straightforward way to assess the changes that are brought by CAVs.  The ultimate goal is to determine whether the changes are of statistical significance.  Figure \ref{fig: impactEvaluationMethod} illustrates the study methodology. First, the human driving behavior is calibrated by multiple sources of data that were collected from the roadway segment of interest. The calibration effort was conducted in \cite{STOLT4}. With a calibrated behavior model, we then treat the car-following model as a black box. The input and the output to the human behavior model is stimuli (local traffic condition) and reactions of HVs, respectively. Then on a collective level, the traffic flow characteristics and vehicle trajectories are analyzed.

\subsection{Simulation of CAV Driving Behavior}
Longitudinal control (or car-following) and lateral control  are the two main components for simulating the behavior of CACC. Numerous car-following models have been proposed. Among them, the Gipps \cite{Gipps1981A}, Wiedemann \cite{Wiedemann1974}, the IDM \cite{Treiber2000} and its variants have been widely used.

\subsubsection{Longitudinal Control}
The Enhanced Intelligent Driver Model (E-IDM) \cite{Kesting2010} is adapt and the longitudinal control model, which is expressed in (\ref{eq: cahCal}).

\begin{equation}
\label{eq: cahCal}
\begin{aligned}
& \ddot{x} = 
\begin{cases}
a[1-(\frac{\dot{x}}{\dot{x_{des}}})^{\delta }- (\frac{s^{*}(\dot{x}, \dot{x}_{lead})}{s_{0}})] & \\ \text{ if } x=  \ddot{x}_{IDM} \geq \ddot{x}_{CAH} \\ 
 (1-c)\ddot{x}_{IDM} + c[\ddot{x}_{CAH} + b \cdot tanh ( \frac{\ddot{x}_{IDM} - \ddot{x}_{CAH}}{b})] & \\ \text{otherwise} 
\end{cases}\\ \quad\\
& s^{*}(\dot{x}, \dot{x}_{lead}) =  s_{0} + \dot{x}T + \frac{\dot{x}(\dot{x} - \dot{x}_{lead})}{2\sqrt{ab}} \\ \quad\\
& \ddot{x}_{CAH} = 
\begin{cases}
\frac{\dot{x}^{2} \cdot \min(\ddot{x}_{lead}, \ddot{x})}{\dot{x}_{lead}^{2}-2x \cdot \min(\ddot{x}_{lead}, \ddot{x})} &  \\ \text{ if }
\dot{x}_{lead} (\dot{x} - \dot{x}_{lead}) \leq -2x \min(\ddot{x}_{lead}, \ddot{x})  \\
\min(\ddot{x}_{lead}, \ddot{x}) - \frac{(\dot{x}-\dot{x}_{lead})^{2} \Theta (\dot{x}- \dot{x}_{lead})}{2x}  & \\ \text {otherwise}
\end{cases} 
\end{aligned}
\end{equation}

where $a$ is the maximum acceleration; $b$ is the desired deceleration; $c$ is the coolness factor; $\delta$ is the free acceleration exponent; $\dot{x}$ is the current speed of the subject vehicle;  $\dot{x}_{des}$ is the desired speed,  $\dot{x}_{lead}$ is the speed of the lead vehicle; $s_{0}$ is the minimal distance; $\ddot{x}$ is the acceleration of the subject vehicle; $\ddot{x}_{lead}$ is the acceleration of the lead vehicle; $\ddot{x}_{IDM}$ is the acceleration calculated by the original IDM model \cite{Treiber2000}; $T$ is the desired time gap; and $\ddot{x}_{CAH}$ is the acceleration calculated by the CAH component as shown in Eq.\ref{eq: cahCal} where $\Theta$ is the Heaviside step function.

\subsubsection{Lateral Control and Coordination}

We assume all the CAVs are equipped with automated longitudinal control. Each CACC vehicle is able to detect the surrounding traffic and discern CAVs from HVs. Three cases are tested:
\begin{enumerate}[label= \roman*)]
\item Base case: case without CAV traffic. This is the baseline for the case study network. The I-66 network has been calibrated.
\item Ad hoc coordination: the CACC system controls only the longitudinal control based on the Enhanced-IDM model. Lateral movement was controlled by the calibrated Wiedenmann model.
\item Local coordination: the CACC control cover both logitudinal and lateral aspects. It was developed in \cite{lee2014mobility}. The longitudinal control of the is replaced with E-IDM and the lateral control is assumed to be done by the automated driving system as well.
\end{enumerate}
A break-down of the cases is shown in Table \ref{table:scenario}. Since the focus of the paper is the near-term deployment of CACC in mixed traffic condition, the MPR is set at a medium-to-low range from 0\% to 40\%.
\begin{table}[H]
\center
\caption{Simulation Case}
\begin{tabular}{l|ll}
\hline
Cases & Longitudinal Control & Lateral Control  \\ \hline
Base & calibrated Weidemann & calibrated Weidemann \\
Ad hoc coordination & E-IDM & calibrated Weidemann\\
Local coordination & E-IDM &  gap acceptance-based \cite{lee2014mobility}\\
\hline
\end{tabular}
\label{table:scenario}
\end{table}

\subsection{Case Study Network}
\begin{figure} [H]
	\centering
	\includegraphics[width=0.5\textwidth]{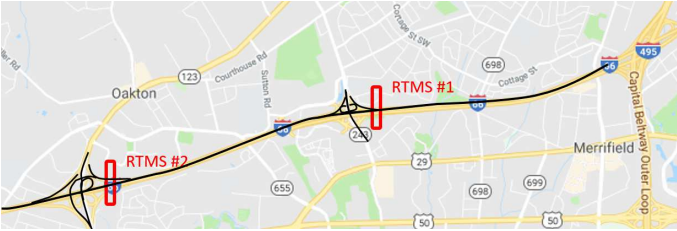}   
	\caption{I-66 Simulation Testbed} 
	\label{fig: aaControlConfig}
\end{figure}

In this study, an 8-km (5-mile) segment (Fig. \ref{fig: aaControlConfig}) of Interstate Highway I-66 outside of the beltway (I-495) of Washington D.C. is used. This freeway segment has recurring congestion during weekdays, specifically in the eastbound direction in the morning and westbound direction in the afternoon. The roadway is with four lanes in each direction. The leftmost lane is an HOV 2+ lane with 1500 vphpl peak volume \cite{lu2014freeway}. Currently, no physical barrier is between the HOV lane and its adjacent GP lane.
The Vissim simulation network is available via the U.S. DOT Open Source Application Development Portal \cite{OSDAP2015}. The calibration was conducted with two independent data sources (i.e., INRIX TMC travel time and RTMS flow data). 

In anticipation of the increase of the traffic demand over time, we assume a 30\% growth in traffic demand from the baseline of the calibrated network. Each deployment scenarios was run five times with different random seeds to factor in the variability of the traffic condition. The duration of the simulation is 3900 s, the first 300 s of which is used to load the network with traffic. No data is collected during this period.
The evaluation of the impact is based on the following assumptions:
\begin{itemize}
\item a low-level vehicle controller for longitudinal (e.g., throttle, brake) and lateral (e.g., steering) control is available. This study only focuses on the tactical driving aspect of vehicle operation.
\item a calibrated driving behavior model in Vissim with real-world data constitutes a good representation of the human driver.
\item vehicle-to-vehicle (V2V) communication is perfect (no interference or packet loss).
\item human drivers do not differentiate CAVs and other HVs as followers
\end{itemize}

\section{Results \& Discussion}
\label{sect:result}
The vehicle trajectory data are collected every 0.5 s, which offers a snapshot of the prevailing traffic condition.  Not only vehicle dynamic data, but also the interaction state including driving state, interacting vehicle, etc. is available.

\subsection{Network Performance}
\begin{figure}[h]
\begin{minipage}[h]{0.5\linewidth}
\centering
\subfloat[productivity]{\frame{\includegraphics[scale=0.35]{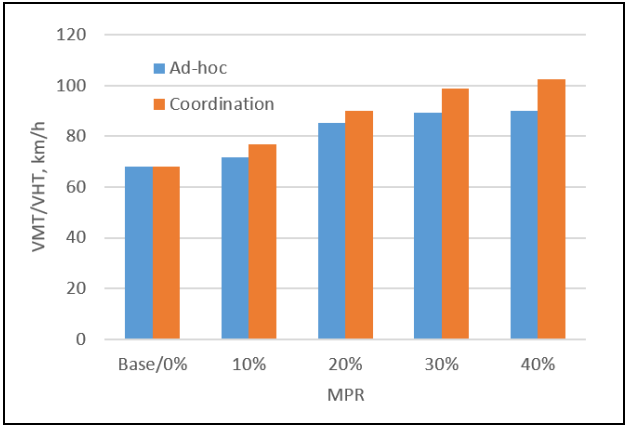}}}
\end{minipage} \\
\begin{minipage}[h]{.5\linewidth}
\centering
\subfloat[throughput]{\frame{\includegraphics[scale=0.35]{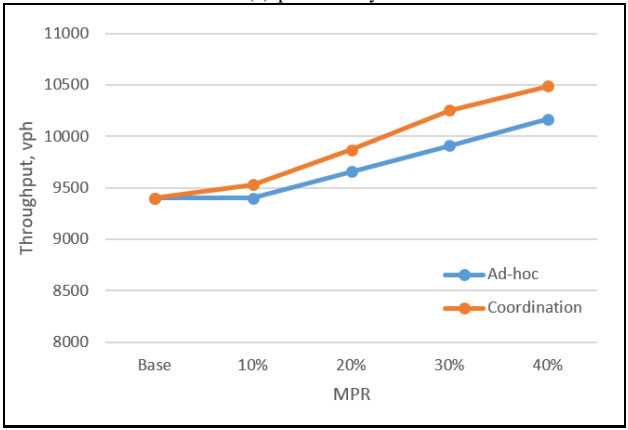}}}
\end{minipage}
\caption{Network Performance}
\label{fig:netPems}
\end{figure}

Fig. \ref{fig:netPems} shows the benefits that CACC bring to the transportation network. Fig. \ref{fig:netPems}(a) shows the ratio of vehicle miles traveled (VMT) and vehicle-hours traveled (VHT).  VMT is  the output of a transportation system, whereas VHT is considered the input to a transportation system. The ratio of VMT and VHT ratio is referred as Q \cite{Caltrain2013aGuide}, which represents the output of a transportation system with the unit value of the input. In short, the higher the value of the Q, the more productive a transportation system is. Both of the ad hoc and local coordination strategies exhibit an increasing trend as the MPR increases. It is notable that the benefits gained by ad hoc coordination show a diminishing increase after 30\% MPR.  In comparison, local coordination displays a linear increasing pattern for the performance gain. 

When it comes to network throughput,  Fig. \ref{fig:netPems}(b) shows that the ad hoc coordination does not increase the network throughput at 10\% MPR: the throughput remains as 9398 vph. After 10\% MPR, the throughput for ad hoc coordination increases with a liner pattern, which matches the underlying operational implication of ad hoc coordination. The throughput reaches the highest 10167 vph at 40\%. With local coordination, additional throughput is observed even at 10\% MPR. Then the slope of the throughput curve is greater at the MPR range between 10\% and 30\% than other tested MPR values.  It also shows that the rate of increase of the throughput decrease after 30\% MPR.  Moreover, the vertical distance indicates the magnitude that the local coordination outperforms the ad hoc coordination at each level of MPR. The greatest difference is observed at 30\% MPR.

\subsection{Hard Braking Observations}
\begin{figure}[h]
\begin{minipage}[h]{\columnwidth}
\centering
\subfloat[hard braking observation CDF]{\includegraphics[scale=0.223]{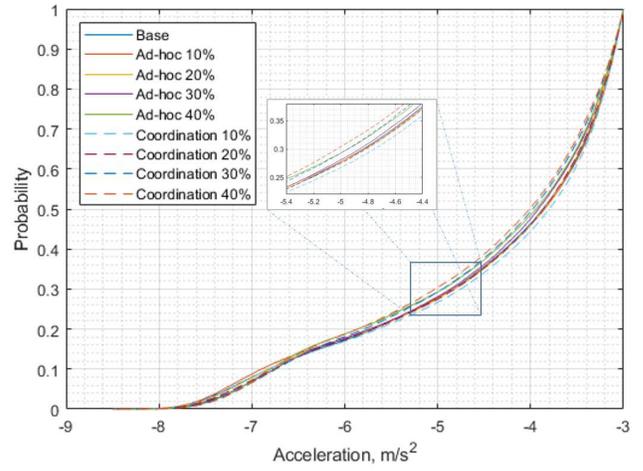}}
\end{minipage}\\
\begin{minipage}[h]{\columnwidth}
\centering
\subfloat[hard braking observation]{\includegraphics[scale=0.45]{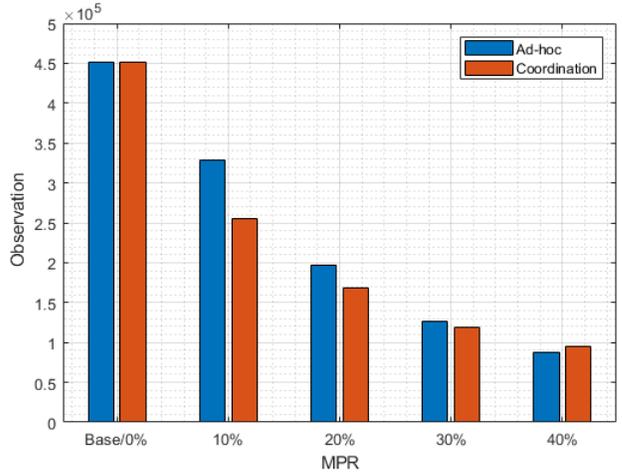}}
\end{minipage}
\caption{Hard braking observation when interacting with HVs}
\label{fig:hdCdfHv}
\end{figure}

Hard braking observation has been used as a safety measure. Abrupt braking is likely an indication of a hazardous traffic situation that drivers respond to \cite{bagdadi2011jerky}. Hard braking observations are recorded when the acceleration of a vehicle is less than -3 $m/s^2$.  Recall our primary focus is the HVs. There are two types of hard braking: the first one occurs when an HV interacts with another HV; whereas the second type occurs when an HV interacts with a CAV.  The hard braking observation for HVs when they interact with other HVs is shown in Fig. \ref{fig:hdCdfHv}(a).  Similar patterns of the cumulative distribution function (CDF) curves for hard braking are observed across the testing scenarios.  Fig. \ref{fig:hdCdfHv}(b) shows the sample size for each scenario. The primary factor for the decreasing trend is the reduction of HVs the traffic stream. The linear trend also infers that the likelihood of hard braking remains at the same level. 

\begin{figure}[h]
\begin{minipage}[h]{\columnwidth}
\centering
\subfloat[hard braking event CDF]{\includegraphics[scale=0.45]{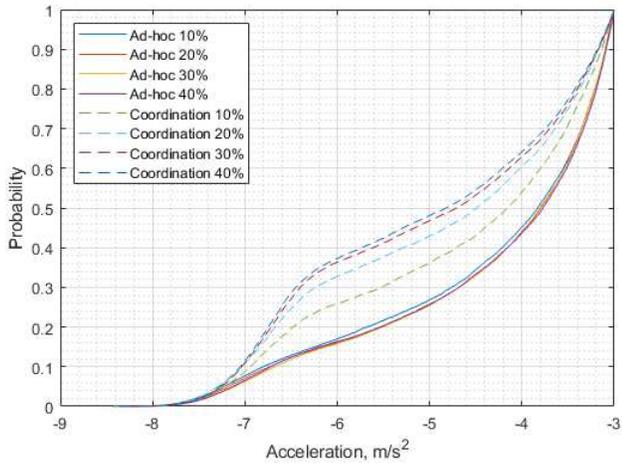}}
\end{minipage}\\
\begin{minipage}[h]{\columnwidth}
\centering
\subfloat[hard braking event sample size]{\includegraphics[scale=0.45]{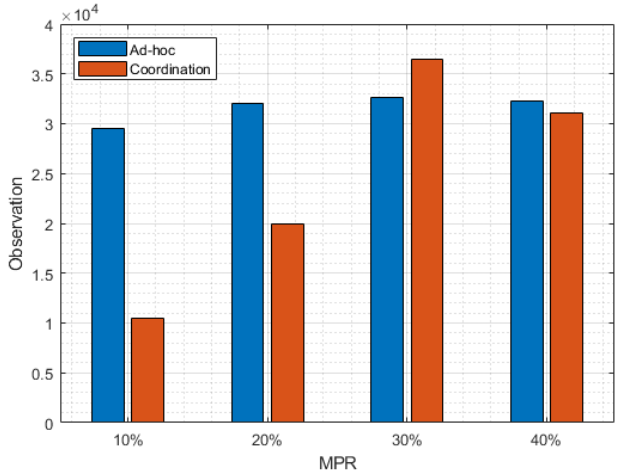}}
\end{minipage}\par
\caption{Hard Braking Event when Interacting with CAVs}
\label{fig:hdCdf}
\end{figure}

Fig. \ref{fig:hdCdf}(a) shows the CDF curves of the hard breaking observations recorded for HVs when they interact with CAVs under each scenario. The CDFs show two distinctive patterns between two coordination strategies. In the ad hoc coordination cases, the CDFs are with similar distributions. On the other hand, the CDF curves of the local coordination are more sensitive to MPR. The probability of hard braking in the range from -6.5 to -3.5 $m/s^2$ drastically increases even at 10\% MPR. The occurrence of hard braking event keeps at the same level in ad hoc coordination; whereas the occurrence of coordination strategy shows an increasing trend until 30\% MPR where the value peaks. 
The sample size is shown in Fig. \ref{fig:hdCdf}(b). Both strategies exhibit an increasing trend until 30 \% MPR, then a declining occurrence after 30\% MPR. With the same amount of CAVs, the hard breaking is more sensitive to MPR in Local coordination then that in ad hoc strategy.

Two-sample Kolmogorov-Smirnov (K-S) test is adapted to further analyze the CDF curves. The K-S test is a powerful tool for testing the hypothesis that whether two random samples have been drawn from the same population \cite{goodman1954kolmogorov}. It is a non-parametric test where no assumption is made regarding the distribution of the variables \cite{young1977proof}. The null hypothesis ($H_0$) of the K-S test is that the comparing two sample sets are from the same continuous distribution. The the two-sample K-S test is conducted for hard braking for each pair of the scenarios at 5\% significance level.  The hypothesis tests show that any pair of the scenarios rejects the null hypothesis and accept the alternative hypothesis that the two samples are not from the same distribution. The $H_0$ is not rejected only in the ad hoc coordination case at 20\% and 30\% MPR.

\subsection{Lane Change Activity}
\begin{figure} [h]
	\centering
	\frame{\includegraphics[width=0.9\columnwidth]{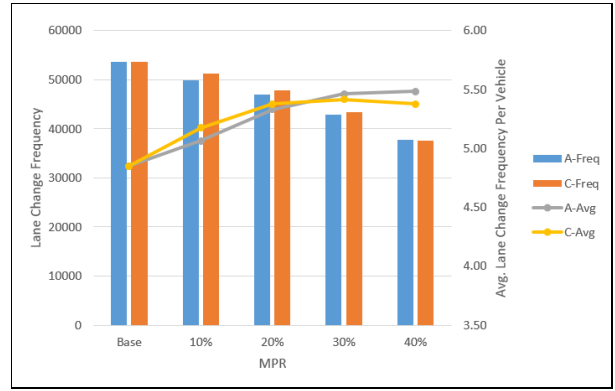}}   
	\caption{Lane Change Activity of HVs} 
	\label{fig: lcFreqHV}
\end{figure}

Fig. \ref{fig: lcFreqHV} shows the accumulative lane change instance recorded at ever 0.5 s for all HVs. The lane change activity of CACC is not recorded as the scope of the paper is confined to HVs.  The lane change activity decrease as the MPR of CACC increase in either of the coordination strategies. Local coordination is marginally higher than the ad hoc one when the MPR is low. At 40\% MPR, they reach the same level of lane change activity. However, recall that the number of HV within the network decreases as the MPR of CACC increase. As such, the average lane change frequency for each HV is calculated and plotted in Fig. \ref{fig: lcFreqHV} as well. The local coordination strategy shows a higher average lane change frequency at 10\% and 20\% MPR. The average lane change frequency peaks at 30\%, then reduced to 5.38 from 5.42. On the contrary, the increasing trend for ah hoc clustering keeps increasing and reach 5.46 and 5.48 at 30\% and 40\%, respectively. 30\% is the break-even point when it comes to average lane change frequency.

In summary, the local coordination strategy outperforms ah hoc coordination strategy across all levels of MPR in terms of network throughput and productivity. However, the induced hard braking for HV should not be overlooked. The distribution of hard braking for HVs changes substantially when they interacting with CAVs under local coordination. Compared to ad hoc coordination at the same MPR,  the probability of the hard braking event in the range of  -7.3 to -6.5 $m/s^2$ is higher. Local coordination causes a higher average lane change frequency for HV
at low MPRs (i.e., 10\% and 20\%). It starts to decrease after reaching 30\% MPR, whereas in the ad hoc strategy the average lane change frequency maintains the increasing trend.

\section{Conclusion}
\label{sect:conclusion}
In this paper, we investigate the two types of coordination strategies for CAV platoon formation. Platoon clustering strategy is a crucial aspect when it comes to deploying CAV in mixed traffic condition in the near term. Agreeing with previous studies, CAV is able to bring benefits to the transportation network even with ad hoc coordination.  Adapting local coordination can further increase the benefits. The impact on HVs is quantified. The distribution of the hard braking observation for HVs, when interacting with CAVs, change substantially with local coordination strategy for platoon formation. In comparison, the distributions for HVs when interacting with other HV show no substantial changes. The average lane change for HVs increases with the presence of CAVs until 30\% MPR is reached.

Future research would be focused on the following areas. First, the lateral control is an underexplored area compared to longitudinal control of CAVs. Further investigation of platoon formation in mixed traffic is desired. Currently, there are only a few platoon coordination algorithms, most of which are rule-based. 
Second, the aggressiveness of the lane change for CAVs when forming a platoon is also an important aspect. As shown, the characteristic of the HVs traffic could be altered. Some of the changes could pose safety concerns for HVs. 
Third, the comparison between clustering strategies should be expanded to include additional local coordination strategies.

\bibliographystyle{IEEEtran}
\bibliography{CAV_HV_Impact_R1}
\end{document}